\documentclass{iaus}
\usepackage{graphicx}

\newcommand{\kms}{\ensuremath{\,\mathrm{km}\,\mathrm{s}^{-1}}}
\newcommand{\kpc}{\ensuremath{\,\mathrm{kpc}}}
\newcommand{\pc}{\ensuremath{\,\mathrm{pc}}}
\newcommand{\cmcube}{\ensuremath{\,\mathrm{cm}^{-3}}}
\newcommand{\Myr}{\ensuremath{\,\mathrm{Myr}}}

\title[Dynamically dominant interstellar magnetic fields] 
{Dynamically dominant magnetic fields in the diffuse interstellar medium}

\author[Fletcher, Korpi, Shukurov]   
{A. Fletcher$^1$, M. Korpi$^2$ \and A. Shukurov$^1$}

\affiliation{$^1$School of Mathematics and Statistics, Newcastle University, NE1 7RU, UK \break email: andrew.fletcher@ncl.ac.uk, anvar.shukurov@ncl.ac.uk\\[\affilskip]
$^2$Observatory, T\"ahtitorninm\"aki (PO Box 14), FI-00014 University of Helsinki, Finland, \break email: maarit.korpi@helsinki.fi}

\pubyear{2009}
\volume{259}  
\pagerange{100--100}
\date{"Nov. 30, 2008" and in revised form ??}
\jname{Cosmic Magnetic Fields: From Planets to Stars and Galaxies}
\editors{K.G. Strassmeier, A.G. Kosovichev \& J.E. Beckman, eds.}
\begin{document}

\maketitle

\begin{abstract}
Observations show that magnetic fields in the interstellar medium (ISM) often do not respond to increases in gas density as would be naively expected for a frozen-in field. This may suggest that the magnetic field in the diffuse gas becomes detached from dense clouds as they form. We have investigated this possibility using theoretical estimates, a simple magneto-hydrodynamic model of a flow without mass conservation and numerical simulations of a thermally unstable flow. Our results show that significant magnetic flux can be shed from dense clouds as they form in the diffuse ISM, leaving behind a magnetically dominated diffuse gas.
\end{abstract}

\firstsection 
\section{The problem: regular magnetic field resists strong shocks}
Magnetic fields in the interstellar medium of the Milky Way and other disc galaxies are, to a first approximation, frozen-in to the interstellar gas: put another way, the magnetic Reynolds number of the medium is sufficiently high
that advection dominates diffusion in the magnetic induction equation. One would therefore expect that strong velocity shear, such as that which occurs in barred galaxies, and large-scale shocks, like the density wave induced shocks along spiral arms, would lead to a strengthening of the magnetic field when the large-scale magnetic field is approximately perpendicular to the shear or parallel to the shock, as is commonly observed in barred (\cite{Beck:02}) and normal spiral galaxies (\cite{Beck:96}).

Surprisingly, observations of the barred galaxies NGC 1097 and NGC 1365 (\cite{Beck:05}) show that the polarized radio emission, tracing the \emph{regular} magnetic field, is hardly affected by the velocity shear of about $200 \kms\kpc^{-1}$, increasing by a factor of $1$--$7$ whereas theoretically one would expect an increase by a factor of about $60$. It is important to note that similar calculations are able to accurately predict the observed increase in the \emph{total} radio emission, as a result of compression and shear of the turbulent magnetic field in the bar. In normal spiral galaxies a similar, but less pronounced, discrepancy is observed. For example in M51 the neutral gas density increases on average by a factor of $4$ in the spiral arms, which should produce an increase by a factor of at least $16$ in polarized emission if the magnetic field were frozen-in and depolarization remains constant, whereas on average there is no observed increase in polarized emission in the spiral arms (Fletcher et al. in prep.).    

\section{The solution: dense clouds detach from the regular magnetic field}
\begin{figure}
\centering
\includegraphics[width=0.32\textwidth]{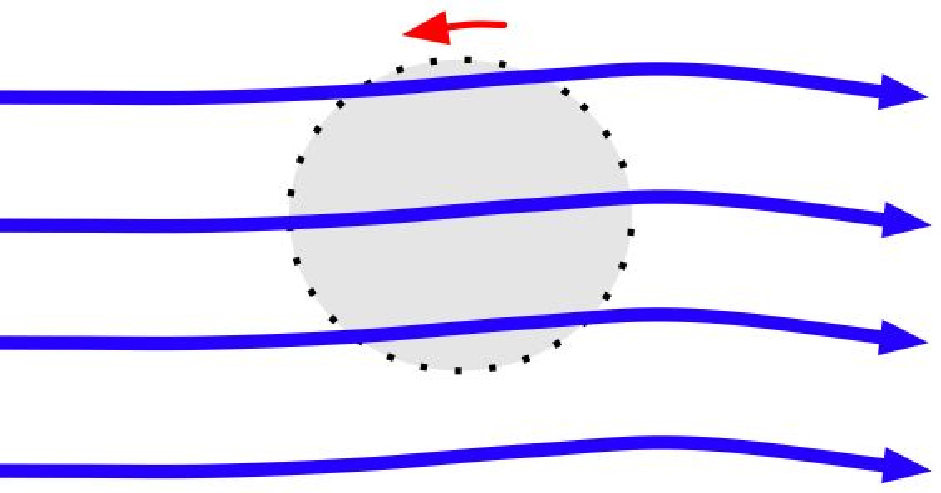}
\includegraphics[width=0.32\textwidth]{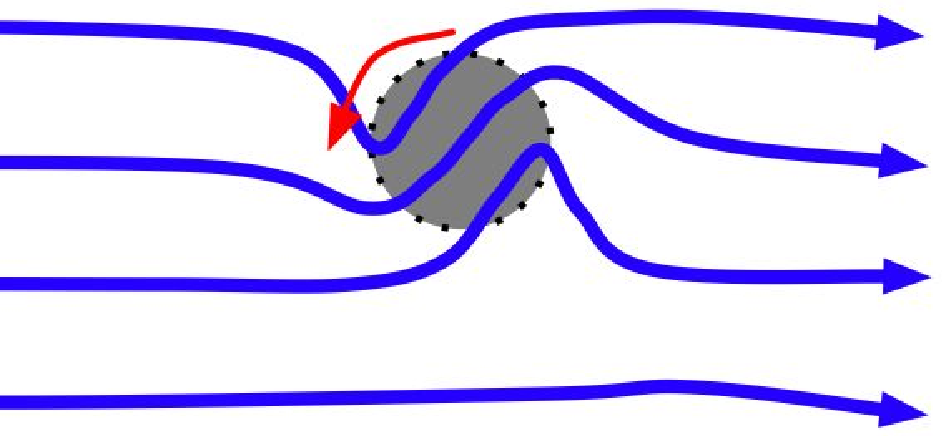}
\includegraphics[width=0.32\textwidth]{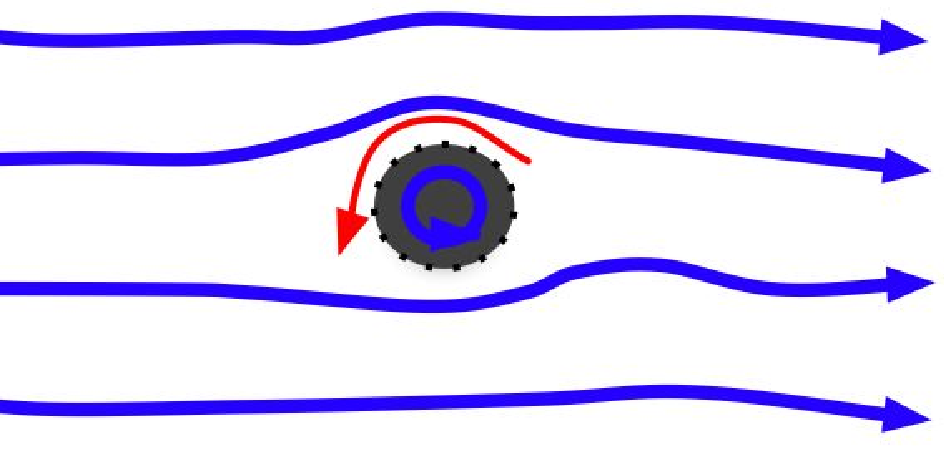}
\caption{Cartoon showing three stages in the condensation of a dense cloud in a diffuse medium threaded by a large-scale magnetic field. As the cloud collapses its rotation velocity increases.}
\label{fig}
\end{figure}

We suggest that the weaker than expected increase in regular magnetic field strength in regions of strong shocks and shear is due to the multi-phase nature of the ISM. The regions in bars and spiral arms where the ISM density increases are also the regions where dense clouds of molecular gas ($n\gtrsim 100\cmcube$) are rapidly formed from diffuse interstellar gas ($n\lesssim 1\cmcube$). As clouds collapse in the turbulent flow, they will rotate faster, winding up the magnetic field lines that thread the cloud. The clouds can then become detached from the background magnetic field in the diffuse ISM by flux expulsion (\cite{Weiss:66}) via ambipolar diffusion (\cite{Mestel:84}) or magnetic reconnection. The cartoon in Figure~\ref{fig} shows the basic principle.

The number of rotations a cloud makes can be bracketed by the following limits. Assuming angular momentum conservation and turbulent flow, an initial density of $1\cmcube$ in a proto-cloud of radius $r=10\pc$ rotating at $v=v_0(r/l_0)^{1/3}\approx 5\kms$, where $v_0=10\kms$ and $l_0=100\pc$ are typical values for the largest turbulent eddies, a cloud will rotate about $40$ times in a typical lifetime of $3$-$5\Myr$ or $100$ times in a formation time of $12$-$20\Myr$. On the other hand, observations of \emph{magnetically braked} mature clouds suggest about $2$ rotations in the formation time (\cite{Bodenheimer:95}). These estimates suggest that sufficient rotations for magnetic reconnection to occur are possible and the model is worth examining in more detail.

We have carried out numerical simulations, in 2D and 3D, of clouds forming via the thermal instability in the diffuse, magnetized ISM. We measure the fraction of magnetic flux threading gas at different densities once the instability has fully developed (after approximately $60\Myr$). The simulations show that magnetic flux is detached from the forming clouds, with the rate of detachment dependent on the magnetic diffusivity. Furthermore, when the perturbations that trigger cloud formation have non-zero rotation the critical density at which flux detachment becomes important scales inversely with the rotation rate. 

If the dense clouds become detached from the regular magnetic field in the diffuse ISM, then the mass-to-flux ratio will decrease for the diffuse gas. Thus the regular magnetic field becomes more important in the dynamics of the diffuse gas and eventually will be strong enough to resist shearing and shocks in the diffuse ISM. This dynamical importance will continue until the dense clouds are dissipated (for example by star formation) and their remaining gas reloads the regular magnetic field in the diffuse ISM.

\end{document}